\documentclass[nodraft,nohyper,notoc]{JHEP} 
\usepackage{epsfig}

\newcommand{\lesim}{\stackrel{\scriptscriptstyle<}{\scriptscriptstyle\sim}}

\title{Probing TeV-scale gauge unification by hadronic collisions}

\author{C. Bal\'azs\thanks{Present address: 
        Department of Physics, Florida State University, 
        Tallahassee, FL 32306, USA}\\
        LPNHE Paris, Universit\'e Paris 7, and University of Hawaii \\
        E-mail: \email{balazs@phys.hawaii.edu}}

\author{B. Laforge \\
        LPNHE Paris, Universit\'es Paris 6-7, and IN2P3-CNRS\\
        E-mail: \email{laforge@in2p3.fr}}

\preprint{\hepph{0110217}\\     
  LPNHE 2001-06\\               
  UH-511-991-01}

\abstract{
Grand unified theories (GUTs) and extra dimensions are potential ingredients 
of the new physics that may resolve various outstanding problems of the
Standard Model.
If the inverse size of (one of) the extra dimension(s) is smaller than the GUT
scale and standard gauge bosons are allowed to propagate in the bulk then, 
among other consequences, the evolution of the gauge couplings deviates from
the usual logarithmic running somewhat below and between these two scales.
In this work we show that if the compactification scale is the order of 10 TeV,
then this modified running may be observable at the CERN Large Hadron Collider 
in the dijet invariant mass distribution.
We also demonstrate that dijets are highly sensitive to the renormalization 
effects of the extra dimensions, and are potential tools for determining the 
number of dimensions and the value of the compactification scale.
}

\begin{document} 

\maketitle 

\section{Introduction}


The goal of theoretical particle physics is to discover a unified theory 
of matter and interactions. String theory appears to be a candidate for such 
a theory, since various string modes can represent the matter, and their 
unique interaction gives rise to all the known forces, gauge and gravitational 
alike \cite{Green:1987sp}. The consistent formulation of superstring theory 
requires ten space-time dimensions to achieve gauge (and gravitational) anomaly 
cancellations \cite{Green:1984sg}, which might be an indication that at the 
fundamental level space-time might have more than four dimensions.


Recently, revolutionary advances have been made in understanding the structure 
of string theory: the discovery of branes \cite{Polchinski:1995mt}, dualities 
\cite{Hull:1995ys,Witten:1995ex}, M-theory \cite{Horava:1996ma}, and the 
AdS-CFT conjectures \cite{Maldacena:1998re}. These developments inspired the 
study of the phenomenological aspects of effective field theory models with low 
fundamental scale(s) and additional space-time dimensions 
\cite{ArkaniHamed:1998rs,Dienes:1998vh,Dienes:1999vg,Randall:1999ee,%
Randall:1999vf}, 
in search for the solution of the gauge hierarchy and the cosmological constant
problems. These models also present a fresh way to interpret other problems,
such as the problem of symmetry breaking. 
These works suggest that it might be possible to formulate string 
theories with a fundamental scale close to the weak scale \cite{Lykken:1996fj}. 
If these weak-scale strings are realized in Nature, then near-future particle 
accelerators may discover stringy phenomena like winding modes or Kaluza-Klein 
(KK) excitations.


In the field theory framework, gauge and Yukawa unification has proven to be an 
attractive assumption to economically explain the diverse features of matter and 
interactions, as for example the large number of parameters in the Minimal 
Supersymmetric Standard Model (MSSM). The advantages of unification and extra 
dimensions can be combined to solve problems of the Standard Model, the MSSM, and 
problems of grand unified theories (GUTs) formulated in four dimensions 
\cite{Kawamura:2001ev,Altarelli:2001qj,Hall:2001xr}. This picture is even more 
attractive considering that the unification and compactification scales can be 
lowered close to the weak scale, naturally avoiding the hierarchy between them.

In the weak-scale compactification scenario, when standard gauge bosons are 
allowed to propagate in the extra dimensions, precision electroweak measurements 
constrain the masses of the KK excitations of the gauge bosons. Global fits to 
electroweak observables provide lower bounds on the inverse compactification 
scale, $1/R_C$, which are generally in the 2-5 TeV range 
\cite{Delgado:2000sv,Rizzo:2000br}. Within this model, the phenomenology of the 
virtual and real production of the KK excitations of the gauge bosons 
at various present and future colliders was recently examined by several 
authors \cite{Antoniadis:1999bq,Nath:1999mw,Dicus:2000hm}. The typical 
constraint on the compactification scale for the CERN Large Hadron Collider (LHC) 
with 100$fb^{-1}$ of 
luminosity is $M_C = 1/R_C \le 6$ TeV \cite{Nath:1999mw}. Stronger constraints 
can be obtained in models with specific assumptions \cite{Accomando:2000sj}.


It is important to note that the earlier phenomenology work does not examine
gauge unification together with the TeV-scale extra-dimensions.  In this paper
we attempt to do so.  The general idea is that, due to the effect of the
extra dimensions, the evolution of the gauge couplings is
modified by power law terms \cite{Dienes:1998vh,Dienes:1999vg}.  This modified
running can potentially be detected in processes which depend sensitively on one
of the gauge couplings.  We illustrate this using dijet production at the
Fermilab Tevatron and at the LHC.  This cross section, at the lowest
order, is proportional to $\alpha_S^2$.  We calculate the cross section,
including NLO QCD corrections, within the Standard Model, assuming that new 
space dimensions open up in the $O(10$ TeV$)$ energy range and gluons can
propagate in these new dimensions.  Then we compare this to the result
of standard QCD, evaluating the statistical significance of the deviation.
In this scenario, we find that in the dijet channel the LHC discovers a common
compactification scale up to $M_C = $5-10 TeV at 5$\sigma$, depending on the 
treatment of KK thresholds as discussed below.  The Tevatron run II should be 
able to discover a 1 TeV compactification scale in the optimistic case.  
We emphasize that these results can be generalized for other processes, 
different colliders, and other models with low scale gauge unification.

\section{TeV-scale gauge unification}

In this section, we outline the theoretical framework that we adopted.
Motivated by the facts listed in the Introduction, we assume the existence of
a higher (than 4) dimensional underlying theory in which all the fundamental
scales are close to the weak scale and the standard gauge bosons propagate in
the extra dimensions.  In particular, gauge and Yukawa unifications happen
somewhat above the TeV energy range at $M_{GUT}$.  In the meantime, the
compactification scale is between the two scales: $M_{Z^0} \lesim M_C = 1/R_C
\lesim M_{GUT}$.  A working example of this is the model outlined in Refs.
\cite{Dienes:1998vh,Dienes:1999vg}.  In this case, the underlying theory has
3+$\delta$ independent space directions and one time dimension.  It is assumed 
that the $\delta$ additional space dimensions compactify on circles with a 
common radius $R_C = O(0.1$ TeV$^{-1}$).  
The known fermions are confined to the observed 4
space-time dimensions, while the known gauge bosons (especially the gluons) and
possibly existing Higgs bosons propagate in the full space (4+$\delta$
dimensions).  If low energy supersymmetry exists, then one can assume that some
of the additional chiral families access the extra dimensions.  In this work we
assume that this does not happen\footnote{In the notation of
\cite{Dienes:1998vh,Dienes:1999vg}, this means that the number $\eta$ of chiral
fermions propagating in the bulk is set to zero.}.


Following \cite{Dienes:1998vh}, we assume that below the compactification 
scale (in the TeV energy range) the theory can be well approximated by a field 
theory formulated in a 4-dimensional space-time. In Refs. 
\cite{Dienes:1998vh,Dienes:1999vg} this theory is assumed to be the MSSM, 
supplemented with the KK excitations of its non-chiral sector. 
It is shown that the presence 
of these KK excitations affects the renormalization evolution of the 
gauge couplings. This effect is quantified by power-law type corrections to the 
usual logarithmic scale dependence. In the scenario of \cite{Dienes:1998vh}, at 
the lowest order, the scale dependence of the gauge couplings is given by
\begin{eqnarray}
       \alpha_i^{-1}(\mu) = \alpha_i^{-1}(\mu_0) -
            {b_i-\tilde b_i\over 2\pi}\,\ln{\mu \over \mu_0} 
          -~{\tilde b_i\over 4\pi}\,
             \int_{r\mu^{-2}}^{r\mu_0^{-2}} {dt\over t} \,
     \left[ \vartheta_3\left( {it\over \pi R^2} \right) \right]^\delta.
\label{KKresult}
\end{eqnarray}
where $i = 1,2,3$ labels the gauge groups of the MSSM, and the coefficients of
the usual one loop beta functions\footnote{The difference of the gauge evolutions
between the Standard Model and the MSSM is negligible compared to the effect of
an O(TeV) size extra dimension in the $\mu$ = 1-10 TeV range.  Keeping in accord
with  Refs. \cite{Dienes:1998vh,Dienes:1999vg}, we use the MSSM beta functions.}
\begin{eqnarray}
        (b_1,b_2,b_3) = (33/5,1,-3)
\end{eqnarray}
are supplemented by new contributions from the properly supersymmetrized 
KK towers
\begin{eqnarray}
      (\tilde b_1,\tilde b_2,\tilde b_3) = (3/5, -3, -6) + \eta \; (4, 4, 4)
\end{eqnarray}
(where, for simplicity, we set $\eta = 0$). In the last term of Eq.
(\ref{KKresult}) $\vartheta_3$ denotes the elliptic Jacobi function and 
\begin{eqnarray}
      r = \pi \,(X_\delta)^{-2/\delta} ~~~ {\rm with} ~~~ 
      X_\delta = {2 \pi^{\delta/2} \over \delta \Gamma(\delta/2)}.
\end{eqnarray}
We note that in Refs. \cite{Dienes:1998vh,Dienes:1999vg} an approximate 
expression is used to calculate the running of the couplings, but in our work we 
use the exact formula (\ref{KKresult}). 

The power-law term in Eq. (\ref{KKresult}) accelerates the running of the gauge 
couplings and makes them meet earlier than the usual unification scale of
$2*10^{16}$ GeV. In particular, for $\eta = 0$, the strong coupling decreases 
faster than what the logarithmic running describes. This deviation 
from the standard evolution is highly enhanced for low compactification scales, 
as illustrated by Fig. 1 of Ref. \cite{Dienes:1998vh}. We found that, using 
Eq. (\ref{KKresult}), the strong coupling decreases by 25\% at 10 TeV for 
$M_C = 10 $ TeV and $\delta = 2$. 


Finally, we point out that in Ref. \cite{Dienes:1998vh} the matching of the
asymptotic regions of the evolution below and above of the KK mass thresholds is
approximate.  In \cite{Dienes:1998vh} it is suggested that above the
compactification scale Eq. (\ref{KKresult}) is used, while below it the same
with vanishing $\tilde b_i$.  This approximation completely neglects the
width of the KK states.  On the other hand, in our case this width is not
negligible.  For KK excitations of gluons the width is given by
\begin{eqnarray}
      \Gamma_n = 2 \alpha_S(Q) m_n,
\end{eqnarray}
where $m_n = n/R_C$ is the mass of the resonance.  Even with a reduced value of
$\alpha_S$, a 10 TeV resonance has a width of about 2 TeV, which is comparable to
the mass.  Thus, a step-function style matching at the threshold is 
pessimistic, because it underestimates the deviation of the running below 
the resonance thresholds.

The matching at the KK thresholds is discussed in detail in 
Ref. \cite{Masip:2000xy}, and is beyond the scope of this work.  
To demonstrate our point, and for simplicity, we use the pessimistic matching 
prescription of Ref. \cite{Dienes:1998vh}. 
Meanwhile we define an optimistic prescription, which somewhat overestimates 
the KK width effect, by equating $\mu_0$ with the $Z^0$ mass in 
Eq. (\ref{KKresult}).  Neither of these prescriptions is correct, but they
can be viewed as extrema of the exact treatment.

\section{Effect on the hadronic dijet production}

In this section we examine the sensitivity of the near future hadronic
accelerators to the effect of the extra dimensions on the running of the strong
coupling.  In order to detect the possible deviation from the standard evolution
of the gauge couplings, we have to select processes which are highly sensitive
to one of the couplings and can be precisely measured at the near future
colliders.  Dijet production is such a process since, at leading order,
it depends on the square of the strong coupling constant $\alpha_S$, and is
independent from the other couplings\footnote{This is important, since the
accelerated running of the gauge couplings can partially cancel if
the process equally depends on more than one of them.}.  Moreover, the dijet
final state can be fully reconstructed experimentally and can be used to
determine the energy $\hat{s}$ that entered into the hard partonic subprocess.
This energy is the virtuality of the particle exchange in the $s$--channel, and
also the scale at which the coupling constant involved in the subprocess should
be evaluated.  Finally, the expected rate of dijet production is relatively high
both at the Tevatron run II and at the LHC.  At the latter, for example, the
production cross section even at 4 TeV jet-pair invariant mass ($M_{jj}$) is
about 0.3 fb \cite{atlastdr:1999fr}.
We will show that this event rate allows for a good statistical discrimination.

To quantify the difference between the standard expectation and the one with the
modified running we define the statistical significance
\begin{eqnarray}
S = \frac{|N_{SM} - N_{XD}|}{\sqrt{N_{SM}}}
\label{Eq:DefS}
\end{eqnarray}
where
\begin{eqnarray}
N_{SM} = \mathcal{L} \int^{M_{max}}_{M_{min}} dM_{jj} \frac{d\sigma_{SM}}{dM_{jj}}
\end{eqnarray}
is the number of dijet events produced with invariant mass higher than $M_{min}$
assuming the standard running of gauge couplings, $M_{max}$ is the maximal dijet
energy at the given collider, and $\mathcal{L}$ is the integrated luminosity of
the experiment at hand.\footnote{In our numerical calculations we use 100 $fb^-1$
for the LHC and 10 $fb^-1$ for the Tevatron.}  
The number of dijet events is calculated using the
modified running, $N_{XD}$, defined similarly to $N_{SM}$ with the standard
cross section ${d\sigma_{SM}}/{dM_{jj}}$ replaced by the extra-dimensional one
taking into account the modified running of $\alpha_S$.

The bulk of the dijets at the Tevatron and the LHC is produced by the standard
model $q{\bar q} \to q{\bar q},gg$, the $qg,{\bar q}g \to qg,{\bar q}g$ and the
$gg \to q{\bar q},gg$ processes.  In the context of TeV-scale extra dimensional
models with gauge bosons propagating in the bulk, dijet production at the LHC
was examined in \cite{Antoniadis:1999bq} and \cite{Dicus:2000hm}.  These works
study the effect of the KK excitations on the rate without modifying the running 
of the strong coupling.  In these studies it was shown that KK excitations of 
the gluon contribute a significant portion and increase the rate while also 
changing the shape of the dijet cross section.  But in a model where $\alpha_S$ 
decreases significantly at scales around 10 TeV due to the effect of the extra 
dimensions, the change of the coupling partially counter-balances the effect of 
the KK excitations leading to a less conclusive signal for the higher 
compactification scales.
    
On the other hand, in \cite{Dicus:2000hm} it was also found that the high
mass gluonic KK excitations tend to decay into dijets with very high
transverse momenta, while the standard model background has lower jet
$p_T$. We use this fact to disentangle the competing effects of the
KK-excitations and the running of the strong coupling.
Fig. 3 of Ref. \cite{Dicus:2000hm} shows that for compactification scales
up to 10 TeV the KK contribution to the cross section is below 1\% if $p_T
\lesim 3$ TeV, and at most a few percent if $p_T = 5$ TeV. That is, for
$1/R_C = 10$ TeV, the KK contribution is negligible up to $M \sim 2 p_T
\lesim 6$ TeV, and it is a few percent up to $M = 10$ TeV.  This conclusion is
further strengthened by Fig. 4 of Ref. \cite{Dicus:2000hm}, which shows that
if events are selected such that the minimal jet $p_T$ is about
0.5 TeV, then the KK contribution is less than 1\% for all the
compactification scales relevant in this work.
For this reason, in the rest of this work we confine ourselves to the study
of jets with 560 GeV $< p_T <$ 5 TeV. With this cut in place, the effect
of the extra dimensions on the strong coupling can be observed unbiased by
the KK excitations.

\FIGURE[t]{
  \epsfig{file=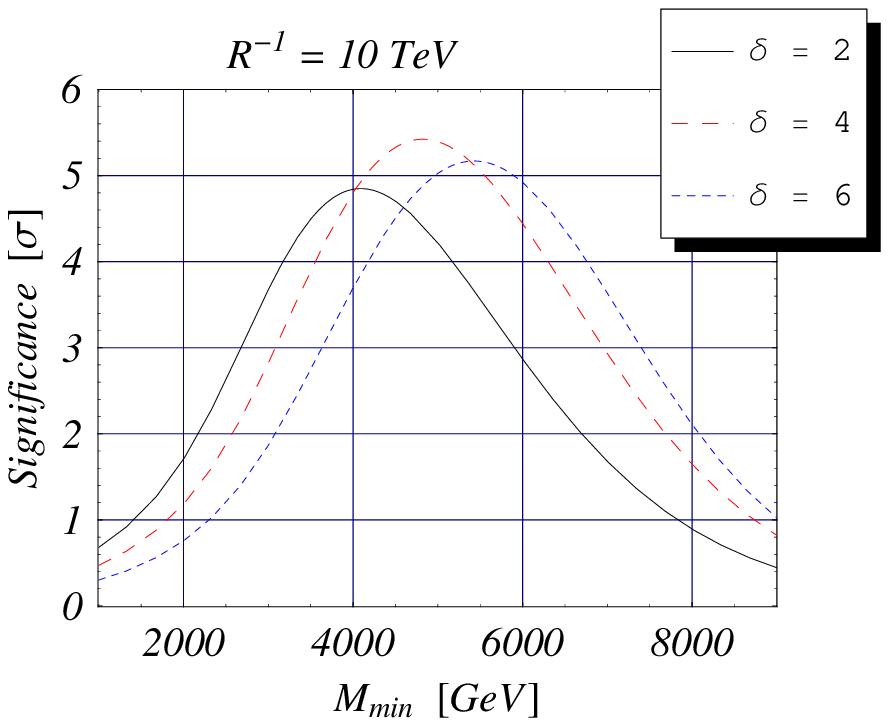,width=12cm} 
    \caption{The statistical significance (in units of $\sigma$'s)
             of the deviation from the Standard Model,
             as the function of the minimal dijet mass $M_{min}$ at the LHC, 
             for different numbers of dimensions and for a compactification 
             scale of 10 TeV. }%
    \label{FigSignificance1}}
In Fig. \ref{FigSignificance1} we show the significance (\ref{Eq:DefS}) as
the function of the minimal dijet mass $M_{min}$ at the LHC for the
optimistic matching prescription.  The qualitative shape of the curves is
easy to understand.  For low $M_{min}$ the small deviation from the standard
running results in a low significance.  For high $M_{min}$ the statistical 
error of the sample increases, which diminishes the significance.  There is an
optimal $M_{min}$ around 5 TeV, where the significance is maximal.

From Fig. \ref{FigSignificance1} we observe that, based on the modified
running of $\alpha_S$, in dijet production the LHC can discover
compactification scales up to 10 TeV at 5$\sigma$ almost independently
from the number of extra dimensions. For the pessimistic matching the
discovery reach is reduced to 5 TeV at 5$\sigma$.  For the Tevatron run II,
based on the projected numbers in Ref. \cite{D0runI:1998}, we obtain a 1
TeV discovery reach at 5$\sigma$ for the optimistic matching scenario.  Fig.
\ref{FigSignificance1} also shows that the peak position of the statistical
significance is sensitive to the number of extra dimensions $\delta$,
which may serve as dimensiometer if $S$ is measured precisely enough as
the function of $M_{min}$.

\FIGURE[t]{ 
    \epsfig{file=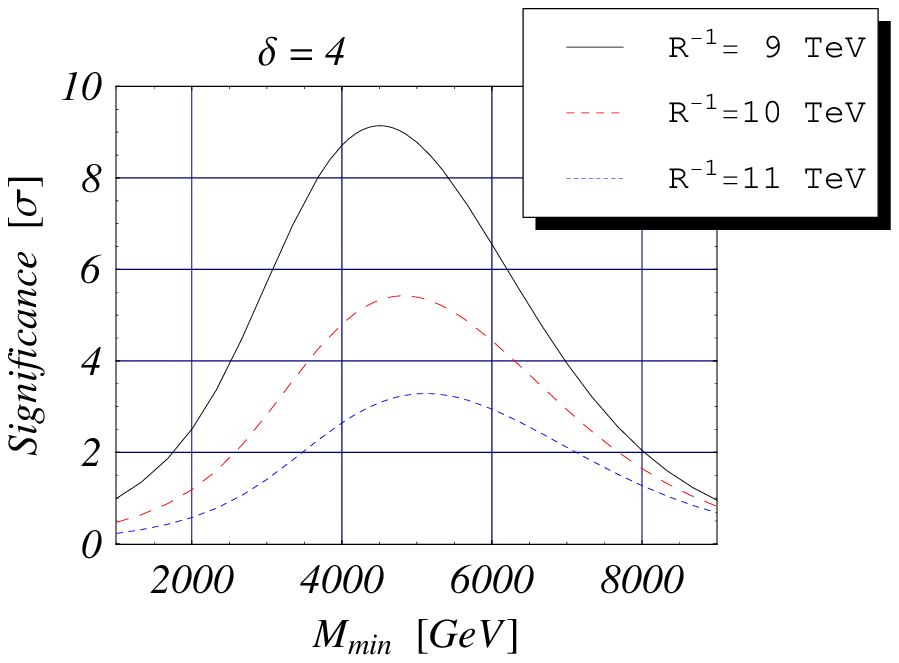,width=12cm} 
    \caption{The statistical significance (in units of $\sigma$'s)
             of the deviation from the Standard Model,
             as the function of the minimal dijet mass $M_{min}$ at the LHC, 
             for different compactification scales and for 4 extra dimensions. }%
    \label{FigSignificance2}}
Fig. \ref{FigSignificance2} shows that the value of the compactification
scale is well correlated with the maximum achievable significance at the
LHC.  This is shown for the optimistic matching scenario, but the
pessimistic one is qualitatively the same.  From the two figures it is clear
that the overall shape of the significance is sensitive to the
compactification scale and the number of extra-dimensions, thus a global
fit to this variable can determine both of these quantities
simultaneously.

Our results also imply that calculations that are performed in a similar 
framework to ours, with compactification scale close to the weak scale and 
(some of) the gauge bosons propagating in the bulk, have to account for the 
modified running of the gauge couplings.  In particular, couplings of KK-%
excitations of gauge bosons to matter are modified significantly at $O$(10 TeV), 
which changes earlier conclusions on their discovery limits.  Also, the modified 
running of the strong coupling might effect the evolution of the partonic 
distribution functions of the proton which are utilized in processes with 
hadronic initial state. These effects are outside of the scope of this work and 
should be investigated elsewhere.

\section{Conclusions}

%
In this work, we examined a scenario in which extra dimensions open up close
above the weak scale, not much below a possible unification scale, 
standard gauge bosons propagating in them.
In this case, the renormalization evolution of the gauge couplings deviates
from the standard running already somewhat above the weak scale.
%
%
We showed that this deviation is measurable at the LHC at 5$\sigma$ in the
dijet channel, provided that the compactification scale is 5-10 TeV$^{-1}$, 
exact values depending on the details of the treatment of the KK thresholds.
The Tevatron run II can observe the extra dimensions up to a maximal scale 
of 1 TeV.
We also demonstrated that dijets are highly sensitive to the 
renormalization effects of the extra dimensions, and are potential tools for 
the extraction of the number of dimensions and the value of the 
compactification scale.

\acknowledgments

The authors are grateful to E. Dudas for useful comments on the manuscript.
We thank the University Denis Diderot which made the collaboration between the 
two authors possible. C.B. thanks LPNHE Paris for its hospitality, and was 
supported in part by the DOE under grant DE-FG-03-94ER40833.


\end{document}